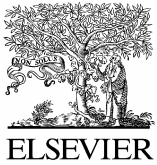

# IceCube and KM3NeT: Lessons and Relations

## Christian Spiering

*DESY, Platanenallee 6, Zeuthen, 15738 Germany*

**Abstract**

This talk presents conclusions for KM3NeT which may be drawn from latest IceCube results and from optimization studies of the IceCube configuration. It discusses possible coordinated efforts between IceCube and KM3NeT (or, for the time being, IceCube and ANTARES). Finally, it lists ideas for formal relations between neutrino telescopes on the cubic kilometer scale.

## 1. Introduction

The IceCube neutrino telescope at the South Pole is approaching its completion and meanwhile has provided first results from data taken with initial configurations [1]. KM3NeT will act as IceCube's counterpart on the Northern hemisphere, with a sensitivity "substantially exceeding that of all existing neutrino telescopes including IceCube" [2]. KM3NeT is just finishing its design phase [3], but has not yet converged to a final configuration. The cited sensitivity requirement results not only from gamma ray observations and their phenomenological interpretation (see e.g. [4-9]), but also from early IceCube data. The first section of this paper is devoted to the lessons for KM3NeT which can be learned from IceCube optimization studies and from first IceCube results.

Assuming one detector on the Southern hemisphere (IceCube) and one or more detectors at the Northern hemisphere (by now ANTARES [10] and NT200 [11], in the future KM3NeT [12] and GVD [13]), various possibilities for coordinated physics programs and analyses open up. They are discussed in sections 3 and 4.

Finally, section 5 lists formal and procedural possibilities of cooperation between KM3NeT and IceCube.

## 2. IceCube: Lessons for KM3NeT

The IceCube collaboration has published results from the 22-string configuration operated in 2007.

Impressively, previous neutrino point source limits obtained from 7 years operation of full AMANDA [14,15] have been superseded by a factor of 2, operating less than a quarter of full IceCube over only one year. This is a step into new territory. But – alas! – neither these data nor preliminary data from the 40-string configuration operated in 2008 reveal any significant excess over the background of atmospheric neutrinos. After corrections for trial factors, the significances for any of several tempting indications decrease to values much below 3σ – be it for steady point sources, for transient phenomena like AGN flares or Gamma Ray Bursts, or for high-energy excesses in diffuse fluxes. This is no surprise. Rather, it is in accordance with recent estimates that even a full cubic kilometer detector may just "scrape" the discovery region.

It is interesting to look back to the years 1999-2002 when the IceCube configuration was optimized. At that time, flux expectations were certainly slightly more optimistic than today but already much lower than a decade earlier, when in the DUMAND-II proposal estimates for a dozen of galactic and extragalactic objects had been listed, ranging between several events and several ten thousand (!) events per source in a cubic kilometer detector [16]. Actually, by the year 2000 it had become obvious that even a full cubic kilometer detector was not a real "guarantee" for source detections. On the other hand, it was argued, that such a detector would dwarf all neutrino detectors existing before the mid-nineties by a factor of thousand. And almost notoriously, improvements of two or three orders of magnitude in astronomy had led to discoveries of new, unexpected phenomena, irrespective of what theoreticians had predicted before [17]. This in mind, a detector on the scale of about one cubic kilometer was envisaged.

*2.1. Spacing and optimum energy range*

The first question was how to arrange four- to ten-thousand 10-inch phototubes to get the best discovery potential for sources with a generic $E^{-2}$ spectrum. In [18] a large number of configurations was simulated, including arrays with equal spacing between the strings, arrays with the string spacing monotonously increasing from the center to outside and vice versa arrays consisting of clusters of strings with local high density similar to those simulated for KM3NeT [3] and even the two rather exotic configurations shown in Fig.1 as a curiosity.

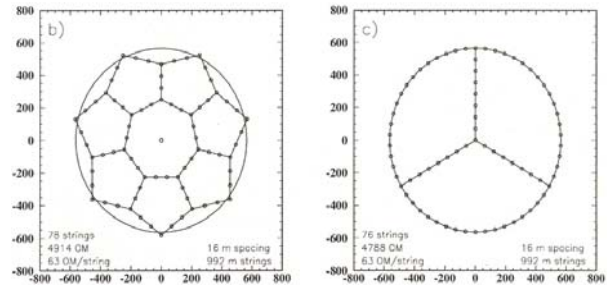

*Fig.1:* Among the ~20 configurations evaluated in early IceCube studies these two have been the most "exotic" ones. For very high energies they performed not worse than standard configurations.

Eventually, the efforts converged to find the optimum value of the string distance assuming *equal spacing*. Fig. 2 shows the number of neutrino events from extraterrestrial sources with a flux $dF/dE = 10^{-6} \cdot E^{-2} \cdot$ GeV cm$^{-2}$ sr$^{-2}$ s$^{-2}$, 1.5 orders of magnitude above the Waxman-Bahcall limit [19], and of atmospheric neutrinos as a function of string spacing. Increasing the spacing from 100 to 180 meters increases the geometrical volume by a factor 3.2, the horizontal area by 80% (assuming fixed vertical spacing between optical modules) and the sensitivity for a 5σ discovery by about 40% [20,21].

One may ask why the actual string spacing was chosen 125 meters instead of 150 m or more. The reluctance was due to several reasons. Firstly, we were more optimistic than today by the reasons mentioned above. For very high absolute fluxes, where events at lower energies are not buried under the atmospheric neutrino background, smaller string spacing makes sense. Secondly, we were afraid that string-to-string calibration by light sources might not work properly over too large distances. Thirdly, the effect of deterioration of the timing information due to the strong light scattering in ice was felt not to be under full control and led to another argument to be "conservative" with respect to spacing.

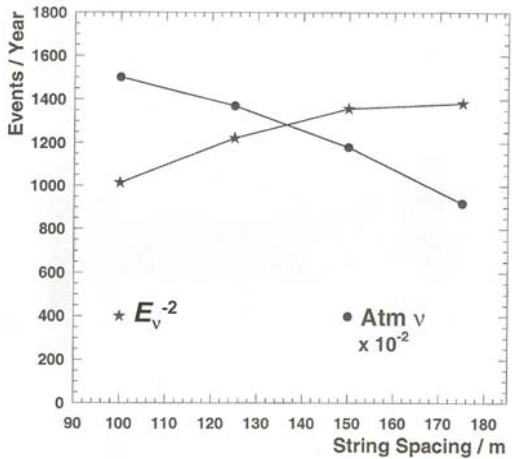

*Fig. 2:* IceCube simulation: Number of neutrino events from extraterrestrial sources with a flux $dF/dE = 10^{-6} \cdot E^{-2} \cdot$ GeV $cm^{-2}\ sr^{-2}\ s^{-2}$ and of atmospheric neutrinos as a function of string spacing [20, 21].

From today's perspective – knowing that the flux is low, knowing that string-to string calibration works extremely well, and knowing that deep ice is better than expected – we very likely would have chosen a spacing of 150-160 meters. It was not a surprise to learn at this conference that corresponding optimizations for KM3NeT result in an optimum string spacing of 130-180 meters (depending on the type of string) [12].

The reason for the preference of large spacing and high-energy thresholds becomes obvious from Figs.3 and 4 (taken from [21]). They show the number of events as a function of energy, with the higher curves corresponding to events passing the cuts for mere rejection of fake events and the lower curves to events passing cuts for reaching the best sensitivity to $E^{-2}$ fluxes. The left parts are for atmospheric neutrinos, the right parts for $E^{-2}$ fluxes with a given normalization, Fig. 3 is for diffuse fluxes and Fig. 4 for point sources. The figures demonstrate that the natural threshold for reaching the ultimate sensitivity for diffuse fluxes is about 100 TeV and that for point sources around one TeV.

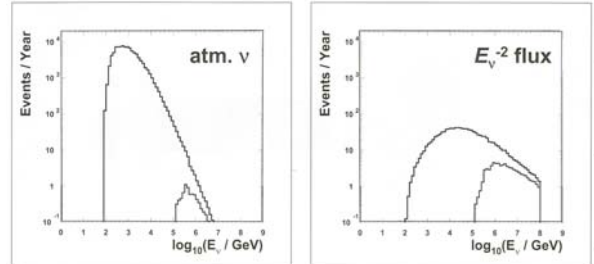

*Fig. 3:* IceCube, diffuse fluxes – Energy spectra for selected atmospheric neutrino events (left) and for selected events from a diffuse extraterrestrial $E^{-2}$ flux (right). The upper curves correspond to cuts which reject fake events from down going muons, the lower curves to cuts which lead to the ultimate sensitivity affordable with IceCube. The cut-off at $10^8$ GeV is due to the limited simulation range.

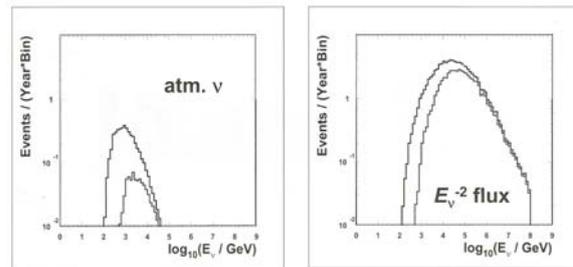

*Fig. 4:* IceCube, point sources – Energy spectra for selected atmospheric neutrino events per search bin (left) and for selected events from a point source with $E^{-2}$ flux (right). The upper curves correspond to cuts which reject fake events from down going muons, the lower curves to cuts which lead to the ultimate sensitivity affordable with IceCube.

One may ask what that means for galactic sources with a steep spectrum or a low-energy cut-off. The answer is nicely illustrated in [8]: If the low-energy flux is below a certain value, inevitably the atmospheric neutrino background dominates and an increase of a densely instrumented volume pays off less and less. Equipping a full cubic kilometer with high density of optical modules is then becoming a waste of resources.

This does not mean that the low-energy region should not be considered. However, for that purpose a much smaller detector will be sufficient, which explores the flux down to the level where it is anyway buried in the background of atmospheric neutrinos. This was one of the rationales behind IceCube's DeepCore which is not much larger than AMANDA (but in better ice, fully surrounded by IceCube and using digital technology). Even with more funding we likely would not triple the volume of DeepCore but would rather invest the money in widely spaced strings around IceCube. The only convincing reason to further increase the volume of DeepCore would be the observation of sources with a low energy cutoff or with a steep spectrum, but strong enough to be identified with DeepCore.

The DeepCore solution, with all dense strings concentrated at one place instead of being spread over several clusters, should have better performance with respect to low-energy cascades (important for oscillations studies). Cascades are important for the search for diffuse extraterrestrial fluxes and for non-standard oscillations of atmospheric neutrinos. Arguably, a too large spacing would have a negative impact on their reconstruction.

### 2.2. How big?

What means "substantially more sensitive than IceCube"? The answer to this question is not independent of what IceCube will see in the next couple of years. The fact that IceCube has not yet identified a single neutrino source means that even with full IceCube one cannot easily expect steady sources with a number of events sufficient for detailed studies (like spectrum measurements). Moreover, predictions for galactic sources suggest that a cubic kilometer detector just scrapes the discovery region (see for instance references [6,7] which suggest optimum thresholds of 5-30 TeV for galactic sources). Under these circumstances "substantially" should be something like the canonical factor 5-10 which is typical for many other next-step projects or upgrades: Auger-North is planned to be about seven times bigger than Auger-South (also sacrificing lower energies!), a conceived 300-ton underground detector for DUSEL would be about six times Super-Kamiokande, and sLHC will have nearly one order of magnitude higher luminosity than LHC.

Part of the gain in sensitivity with respect to IceCube – may be a factor 2 for high energies – will come from the better angular resolution of KM3NeT, but another part can only come from larger geometrical size!

### 2.3. Constraints from diffuse fluxes

The previous section has been about point source searches. IceCube will also search for diffuse high energy excesses from extraterrestrial neutrinos. If IceCube would not see an excess, this would put limits to the number of extragalactic point sources which one reasonably could expect (not to be confused with any "number of events"!). The argument was made in [5]: Contributions to the diffuse flux will come from all over the observable universe, up to a distance $c/H_0$, whereas point sources, with several events per source, will be visible only up to a limited distance of a few hundred Mpc, assuming reasonable maximum luminosities per source. For a homogeneous distribution of extra-galactic sources, one therefore can derive a limit on the number of observable point sources. In [5] the following assumptions are made: a) a homogeneous source density in an Euclidian universe, b) a source luminosity $L_{source}$ "typical" (and similar) for all sources, c) $E^{-2}$ spectrum of sources. Then, assuming an experimental limit $K_{diffuse}$ on the diffuse flux and a sensitivity $C_{point}$ to point sources, the expected number of resolvable extragalactic point sources, $N_s$ is

$$N_s \sim \frac{K_{diffuse} \cdot \sqrt{L_{source}}}{C_{point}^{2/3}}$$

With the present diffuse limit from AMANDA and the expected point source sensitivity of IceCube, one obtains $N_s \sim 1\text{-}10$ [22]. This means that, with the given assumptions, a cubic kilometer detector would have a fair – but not overwhelmingly large – chance to detect extragalactic point sources! If IceCube would push $K_{diffuse}$ twenty times below that of AMANDA, one would have $N_s \sim 0.05\text{-}0.5$, a discouraging small number. Still, with a few individual, very close sources one may circumvent the homogeneity assumption and these could be still observable. Also, point sources with cut-offs below a few hundred TeV would not be covered by the argument above since, in order to obtain the best sensitivity for *diffuse* fluxes, IceCube will place energy cuts at about 100 TeV [20].

Systematic uncertainties influence excesses in diffuse fluxes much more than they do for point source excesses on a sky map. Therefore, whatever effect on diffuse fluxes will IceCube report – an independent confirmation from a technology with different systematics (water instead of ice!) would be important.

### 3. Coordinated physics activities

Coordinated activities include, for instance, common skymap analyses, coordination of alert programs, and definition of common standards in analysis and presentation.

*3.1. Combined analyses*

If neutrino detectors would be sensitive only up to the horizon, at any given time there would be a 25% overlap of the fields of view; moreover, 70% of the IceCube sky would be seen at some moment also by a Mediterranean detector [23]. Actually, however, the overlap is much larger since at high energies the detectors have also some sensitivity above horizon [24] (see Figs. 5 and 6, taken from [25]). This allows a lot of combined analyses:
- Adding statistics to increase the significance of an observation.
- Creation of $4\pi$ full-sky maps.
- In case of a strong source with enough statistics, with a spectrum extending over 3-5 orders of magnitude, one may even think about a neutrino "multi-wavelength" analysis for sources which are seen dominantly above horizon (i.e. at very high energies) by the one, and dominantly below horizon (i.e. at lower energies) by the other detector (see Fig.6).

The minimum requirements for such analyses include, for instance:
- coordinated unblinding procedures,
- coordinated candidate source lists,
- exchange of parameters necessary for point source analyses, like the point spread function or the effective area as a function of energy and angles.

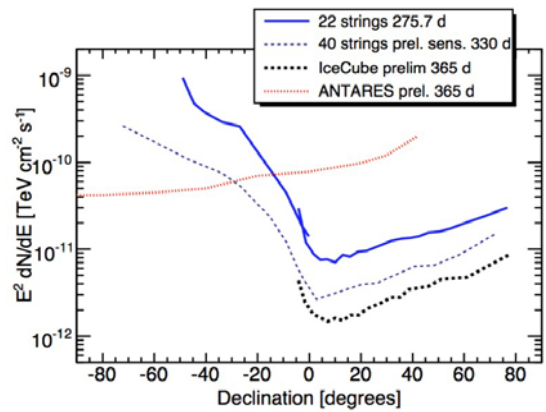

*Fig 5:* *Sensitivity for point sources as a function of declination, obtained from IceCube 22 strings, calculated for 1 year IceCube 40 strings and IceCube 80 strings (1 year) and compared to the expected sensitivity of ANTARES (1 year) [25].*

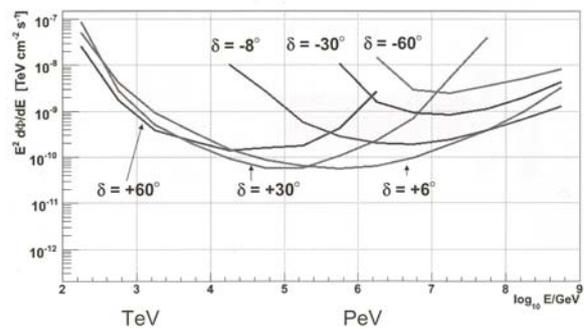

**Fig. 6:** *Differential $5\sigma$ discovery potential as a function of energy and of declination angle (calculated for IceCube 40 string, 1 year). Declinations $\delta > 0°$ correspond to the Northern hemisphere, $\delta < 0°$ to the Southern hemisphere [25].*

*3.2. Coordinated alert programs*

Both IceCube and ANTARES have started, or are preparing, alert programs:
- IceCube prepares two follow-up programs. Within the Neutrino Target of Opportunity program (NToO), the MAGIC telescope will be pointed to one of a few selected sources, if several neutrinos from this source are observed in typical time intervals of hours to days [26]. Another follow-up program will alert robotic optical telescopes (ROTSE) in case of neutrino doublets or triplet from *any* direction, but with typical time windows of seconds to minutes [27]. The first program is tailored to high-states, or flares, of AGN, the second to supernova collapses and GRB.
- ANTARES is already running an optical follow-up program (TAToO) similar that of IceCube using the TAROT robotic telescopes [28].

Other alert programs are being discussed. Clearly, the ratio of signal to background alerts from neutrino telescopes is an issue. Alert programs have to be coordinated worldwide, be it only not to swamp optical/gamma telescopes with an unreasonable number of alerts from neutrino telescopes.

Another idea presented in [23] is that IceCube triggers KM3NeT to take data with lower threshold. If the possibility to lower the threshold on request would be also implemented in IceCube, the process could also work the other way around.

*3.3. Other examples*

Other examples for coordinated efforts are:
- Unification of the style of presentations (How are upper limits calculated and presented? To which models are data compared?),
- Cross check of results from diffuse searches at high energies, which are particularly sensitive to systematic uncertainties – be it excesses possibly related to prompt or extraterrestrial neutrinos, or deficits related to non-standard oscillations and quantum gravity effects [29].
- Confirmation of classes of exotic events, like slow monopoles or Q-balls.

**4. Sharing of software and algorithms**

Sharing of software between AMANDA and ANTARES has a long history, starting with the initial use in ANTARES of simulation programs which had been developed for DUMAND and AMANDA [30]. More recently, in summer 2008, a Memorandum of Understanding between the IceCube and KM3NeT collaborations has been signed on the use of the software framework IceTray [31] (later transformed to KM3Tray and meanwhile to SeaTray [32]). Improvements and debugging will profit from more people using the software. Event generators, air shower parameterizations, reconstruction methods, use of waveforms and basic algorithms (like, already now, the Gulliver fitting code [33]) are other examples for potential cooperation in the field of software.

**5. Formal relations**

There are several levels of formal cooperation:
- *Memoranda of Understanding on specific items.* With the MoU on IceTray software this became already reality.
- *Yearly common meetings.* In September 2009, the IceCube and ANTARES collaborations (inviting also members of the underwater community not part of ANTARES) already had a first common meeting in Berlin, to be followed by a next meeting in Paris in September 2010.
- *Inter-collaboration working groups:* These groups could „synchronize" statistical methods, ways of presentation, simulations etc.
- Forming a Global Network, similar to LIGO/Virgo/GEO including also Baikal/GVD.
- *Forming a Global Neutrino Observatory:* This model would be close to what CTA and the Pierre Auger Observatory have in mind.

The first three forms of common activities are already reality or can easily start just now. The last two will have to wait until at least one of the other next-generation neutrino projects is going to start data taking and then could meet IceCube on equal footing. The best IceCube itself can do to make this reality is to detect a first source: that would be the most efficient boost for the field of high energy neutrino astronomy and for KM3NeT!

**Acknowledgments**

The last three sections of this paper are based on discussions and talks given at the MANTS meeting, (Berlin, September 2009). Special thanks go to Teresa Montaruli, Jürgen Brunner and Chad Finley for helpful information. I also acknowledge discussions with Uli Katz and Tom Gaisser.